\documentclass[preprint,aps,showpacs,nofootinbib,preprintnumbers,amsmath,amssymb]{revtex4-1}
\usepackage{epsfig}
\usepackage{bm}
\usepackage[usenames ,dvipsnames]{xcolor}
\usepackage{slashed}
\usepackage{graphicx,color}

\usepackage{multirow}
\begin{document}
\title{Time-reversal asymmetries and angular distributions in $\Lambda_b \to \Lambda V$}

\author{Chao-Qiang Geng$^{1,2,3}$ and  Chia-Wei Liu$^{1,2}$}
\affiliation{
	$^{1}$School of Fundamental Physics and Mathematical Sciences, Hangzhou Institute for Advanced Study, UCAS, Hangzhou 310024, China \\
	$^{2}$International Centre for Theoretical Physics Asia-Pacific, Beijing/Hangzhou, China \\
	$^{3}$Chongqing University of Posts \& Telecommunications, Chongqing 400065, China
}\date{\today}

\begin{abstract}
We study the spin correlations to probe time-reversal~(T) asymmetries in the decays of $\Lambda_b \to \Lambda V ~(V=\phi, \rho^0, \omega, K^{*0})$. 
The eigenstates of the T-odd operators  are obtained along with definite angular momenta. We obtain the T-odd spin correlations from the complex phases 
among the helicity amplitudes. We give the angular distributions of $\Lambda_b \to \Lambda(\to p \pi^-)V(\to PP')$ and show the corresponding spin correlations, where $P^{(\prime)}$ are the pseudoscalar mesons.  Due to the helicity conservation of the $s$ quark in $\Lambda$, we deduce that the  polarization asymmetries 
of $\Lambda$ are close to $-1$. Since the decay of $\Lambda_b \to \Lambda \phi$ in the standard model~(SM) is dictated by the single weak phase from the product 
of CKM elements, $V_{tb}V_{ts}^*$, the true T and CP asymmetries are suppressed, providing a clean background to test the SM and search for new physics.
 In the factorization approach,  as the helicity amplitudes in the SM share the same complex phase, T-violating effects are absent. Nonetheless, the experimental branching ratio of $Br(\Lambda_b \to \Lambda \phi) = (5.18\pm 1.29 )\times 10^{-6}$ suggests that the nonfactorizable effects or some new physics play an important role. By parametrizing the nonfactorizable contributions with the effective color number, we calculate the branching ratios and direct CP asymmetries. We also explore the possible T-violating effects from new physics.
\end{abstract}
\maketitle
	
\section{Introduction}
 The time-reversal~(T) symmetry demands that  physics shall be invariant under reversing the motions, which corresponds to flipping time $t$ to $-t$ and taking the complex conjugate of the  
quantum states.  
 Its anti-linear property makes T-violation and CP- violation become synonyms, according to the CPT symmetry.
 In two or three-body decays,  T-odd observables can be constructed in the spin correlations among the particles~\cite{ref1,ref2,ref3,ref4,ref5,ref6,ref7,ref8,Li:2010ra,Geng:BtoVV}.
 Since the particle spins could not be directly measured in the current high energy physics experiments, we have to extract their effects through the angular distributions in the cascade decays~\cite{Chiang:1999qn,French0,French1,Hrivnac:1994jx,French2,KornerSM,KornerSM2,KornerSM0,Korner:1992wi,SU3Pseudo,SU3Vec,Cen:2019ims,Li:2021qod,Huang:2021ots}. 
 
There are more than forty naive T-odd observables, which  have been measured in the $B$ meson decays~\cite{pdg}, indicating the angular analyses have been well developed in the experiments. Particularly, the longitudinally polarized fraction, $f_L(B^0 \to \phi K^{*0})\approx 0.5$, measured by  BarBar~\cite{BaBar:2003spf} and Belle~\cite{Chen:2003jfa}, shows that the nonfactorizable (NF) effects  play an important role in the $B$ decays with vector mesons
in the final states. On the other hand,  
 the angular analyses in $\Lambda_b \to J/\psi \Lambda$~\cite{ExpPolarized1,ExpPolarized2,ExpPolarized3} from LHCb
 show that the polarization fractions  
 are consistent with zero. Furthermore, the $\Lambda_b-\overline{\Lambda}_b$ production asymmetry has also been studied~\cite{LHCb:2017slr}.
Recently, the branching ratios of $\Lambda_b \to \Lambda \phi$ and $\Lambda_b \to \Lambda \gamma$\, have been measured to be
$(5.18\pm 1.29)\times10^{-6}$~\cite{LambdabLambdaPhiExp} and $(7.1\pm 1.7)\times10^{-6}$~\cite{LambdabLambdaGammaExp}\,, respectively. However, there have been no complete angular analyses for $\Lambda_b\to \Lambda V (V=\phi, \rho^0,\omega, K^{*0})$.
 
On the theoretical side, to deal with $\Lambda_b$ decays, various approaches have been made~\cite{Dery:2020lbc,Franklin:2020bvb,Han:2021gkl,Roy:2019cky,Roy:2020nyx}. 
The most simple one is the naive factorization~\cite{NaiveFacZhao:2018zcb,NaiveKhodjamirian:2011jp,KornerSM,KornerSM2}, in which the mesons are produced  from weak vertices directly. In particular, the experimental branching ratio of $\Lambda_b \to p \pi ^ -$ can be explained by light-cone QCD sum rules~\cite{NaiveKhodjamirian:2011jp}, and
the experimental results in $\Lambda_b \to J/\psi \Lambda$ are well compatible with those in
 the covariant confined quark model~\cite{KornerSM,ExpPolarized3}.
 In Refs.~\cite{GeneralGeng:2016kjv,GeneralGeng:2021nkl,GeneralHsiao:2017tif,GeneralModifiedBagModel}, the generalized factorization approach has been considered, in which the renormalization dependence of the Wilson coefficients has been absorbed into the effective ones in the next to leading log precision~\cite{EffectiveWilson}. However, the effective color number, found as $N_c\approx 2$ in the  decays of $B$ mesons, has to be included to parametrize the NF effects in the branching ratios.
On the other hand, the NF amplitudes can be calculated systematically based on the QCD factorization~\cite{Zhu:2018jet} and  perturbative QCD~\cite{Lu:2009cm} approaches, in the heavy quark limit. Nevertheless, they suffer large uncertainties in the penguin dominated decays, due to the unknown baryon wave functions. In this work, we adopt the generalized factorization approach to estimate the results in the standard model (SM).
 
We will study T-violation systematically in $\Lambda_b \to \Lambda V$ based on the group theory approach.
 The T-violating observables are often given by the triple vector products asymmetries, read as 
 \begin{equation}~\label{old}
{\cal A}_T = \left[\Gamma ( \hat{T} > 0 )-\Gamma ( \hat{T} < 0  )\right]/\Gamma\,,
 \end{equation}
where $\Gamma$ corresponds to the decay width, $\hat{T}= (\vec{v}_a \times \vec{v}_b) \cdot \vec{v}_c$\,,  and $\vec{v}_i$ represents  either the spin ($\vec{s}$) or 3-momentum ($\vec{p}$) of the particle labeled by $i$ with $i=a$ or $b$ or $c$~\footnote{Since angular momentum is always conserved, the spin of $\Lambda_b$ can be identified as the angular momentum in the final state.}.
Under the T transformation, $\hat{T}$ flips its sign, and therefore ${\cal A}_T$ are naively  T-odd observables.
However, cares must be taken when one applies it on the vector products involving spins.
For instance, $(\vec{s} \times \vec{p}) \cdot \vec{J}$ is  Hermitian, but  $(\vec{s} \times \vec{J}) \cdot \vec{p}$ is {\it not}, with $\vec{J}$  the angular momentum operator. The Hermiticity can be understood by the commutation relation,
\begin{equation}
[(\vec{A}\times \vec{B})_i , J_i] = 0 \,,
\end{equation}
where $\vec{A}$ and $\vec{B}$ are arbitrary vectors. In contrast, the equality, 
\begin{equation}
(\vec{A} \times \vec{J}) \cdot \vec{B} = -(\vec{A} \times \vec{B}) \cdot \vec{J} + 2 i \vec{A} \cdot \vec{B}\,,
\end{equation}
 indicates that $(\vec{s} \times \vec{J}) \cdot \vec{p} $ is {\it not} Hermitian. Hence, it clearly makes no sense  to discuss the case with $\vec{v}_b = \vec{J}$. 
Another issue arises from  an incompatible set of observables,  causing inconsistent in the analyses, which will be discussed carefully.

This paper is organized as  follows.
In Sec.~\MakeUppercase{\romannumeral 2}, we study the angular distributions of $\Lambda_b \to \Lambda(\to p \pi^-) V (\to PP')$\,,  where $P^{(\prime)}$ are the pseudoscalar mesons in the cascade decays. In Sec.~\MakeUppercase{\romannumeral 3}, we define the T-odd observables and identify their effects in
terms of  the angular distributions. In Sec.~\MakeUppercase{\romannumeral 4}, we examine the decays in the SM based on the generalized factorization approach. We also discuss possible right-handed current contributions
to T-violating observables
from new physics.
Finally, we give conclusions in 
Sec.~\MakeUppercase{\romannumeral 5}.

\section{Angular distributions}

The decay distributions in two body decays are often described by the helicity amplitudes~\cite{Jacob:1959at}.
In general, the  decays amplitudes of $\Lambda_b \to \Lambda V$ take the form,
\begin{equation}
\langle \Lambda V; \{ \lambda _f\} | {\cal H}_{eff} | \Lambda_b\rangle\,, 
\end{equation}
where ${\cal H}_{eff}$ is the effective Hamiltonian of weak transitions, and $\lambda_f$ correspond to physical quantities to characterize the states properly. Often, $ \lambda_f$ are chosen as the momenta and spins of the particles, resulting in expanding the amplitudes with the Dirac spinors and polarization vector, given as 
\begin{equation}\label{badEXP}
\xi^{\mu*} \overline{u}_\Lambda \left[
A_1\gamma_\mu +\frac{A_2P_\mu }{m_{\Lambda_b}} - \left(B_1\gamma_\mu + \frac{B_2P_\mu }{m_{\Lambda_b}}\right)\gamma_5
\right]u_{\Lambda_b}\,,
\end{equation}
where  $ A_1,A_2,B_1$ and $B_2  $  are the effective couplings, $\xi^{\mu}$ represents the polarization vector of $V$, $P^\mu=p^\mu _{\Lambda_b}+p^\mu _{\Lambda}$, and $p^\mu_{\Lambda_{(b)}}$ $(u_{\Lambda_{(b)}})$  are the four momenta (Dirac spinors) of $\Lambda_{(b)}$\,.
Expanding the amplitudes in  momenta and spins has the advantage in the dynamical aspect, for that the amplitudes are easier to be parametrized in the calculations, such as the framework of the factorization approach.

On the other hand, from the kinematical consideration, it is more preferable to describe the states in terms of $J^2$ and $J_z$.
Since the values of $J^2$ and $J_z$ are always constrained by the parent particle,  one may eliminate the redundancies caused by the $SO(3)$ rotation group~($SO(3)_R$), which are independent of the dynamical details. 
Nonetheless, $J^2$ and $J_z$ do not specify the states unambiguously.
We can choose a set of commuting $SO(3)_R$ scalars to identify the states further. Concerning the angular distributions in  the sequential decays, the helicites can be a good choice. In the rest frames of $\Lambda$ and $V$, the spins can be read directly.

In the center of the momentum frame of the $\Lambda V$ system, the helicity and 3-momentum states are related as~\cite{Jacob:1959at,GroupTheory} 
\begin{eqnarray}\label{HelicityStates}
|\lambda_1\,, \lambda_2 ; J=1/2, J_z=M\rangle = 
\frac{1}{2\pi}
\int d\Omega |\vec{p}_1, \lambda_1\,, \lambda_2\rangle  e^{iM\phi} d^\frac{1}{2}(\theta) ^{M}\,_{\lambda_1 - \lambda_2}\,,
\end{eqnarray}
where $d^\frac{1}{2}$ stands for  the Wigner d-matrix for $J=1/2$, 
$\lambda_{1(2)}$  represents the helicity of $\Lambda~(V)$, and $\vec{p}_1$ corresponds to the 
3-momentum of $\Lambda$ in the $\Lambda_b$ rest frame.
In Eq.~\eqref{HelicityStates},
the left-hand side is the so-called helicitiy state, while the right-hand one is made of the linear superposition of the 3-momenta. To have a nonzero value in $d^{\frac{1}{2}}$, one must have $ 1/2\ge |\lambda_1 - \lambda_2|$. The helicity states are given as
 \begin{eqnarray}
|a_\pm \rangle = | \pm 1/2 \,, 0\rangle \,,\,\,\,\,\,\,\,|b_\pm \rangle = | \mp 1/2 \,, \mp 1 \rangle\,,
 \end{eqnarray}
where the first and second entries correspond to $\lambda_1$ and $\lambda_2$, respectively, with $J=J_z=1/2$. In this work,  unless explicitly stated,  $J$ and $J_z$ will not be written out explicitly.
Here, $a~(b)$ indicates the vector meson is longitudinally~(transversely) polarized, while the subscripts denote the the angular momenta in the  $\hat{p}_1$ direction, 
read as $\vec{J}\cdot \hat{p}_1 = \lambda_1 - \lambda_2 = \pm 1/2$.
Under the parity transformation, the helicities flip signs, given as 
\begin{equation}\label{Is}
I_s|a(b)_\pm \rangle = |a(b)_\mp \rangle  \,,
\end{equation}
 where $I_s$ is the parity transformation operator.

Consequently,
the decays amplitudes are  given by
\begin{eqnarray}
a_\pm =H_{\pm\frac{1}{2},0 } = \langle a_\pm ; \text{``out''} | {\cal H}_{eff} | \Lambda_b\rangle\,,\nonumber\\
b_\pm =H_{\mp\frac{1}{2},\mp 1 } = \langle b_\pm ; \text{``out''} | {\cal H}_{eff} | \Lambda_b\rangle\,,
\end{eqnarray}
where ``out'' denotes $t\rightarrow \infty$. 
If ${\cal H}_{eff}$ respects the T symmetry, one has that\footnote{
In practice, $J_z$ flip sign under the TR transformation. Nonetheless, since the angular momentum is conserved, we can rotate both sides back  without affecting the amplitudes.
}~\cite{GroupTheory}
\begin{equation}
a(b)_\pm = \langle  \Lambda_b | {\cal H}_{eff} |  a(b)_\pm ;\text{``in''}\rangle\,,
\end{equation}
where ``in'' denotes $t\rightarrow -\infty$\,. Furthermore, if the final state interactions~(FSIs) are absent, we can interchange ``in'' and ``out'' freely, resulting in 
\begin{equation}\label{usualTR}
a(b)_\pm = \langle  \Lambda_b | {\cal H}_{eff} |  a(b)_\pm ;\text{``in''}\rangle = a^*(b^*)_\pm \,.
\end{equation}
As a result, one concludes that $a(b)_\pm$ are real.
This is  a common approach 
in analyzing the T symmetry with complex phases.  

The angular distributions
are parametrized with five angles, $\theta\,,\theta_{1,2}$ and $\phi_{1\,,2}$\,, shown in FIG.~1. In the following, we take $\Lambda_b \to \Lambda (\to p \pi^-) \rho^0 (\to \pi^+ \pi^-)$ as a concrete example.
In this case, $\theta\,, \theta_1\,,$ and $\theta_2$ are defined as the angles between $(\hat{n}_{\Lambda_b}, \vec{p}_\Lambda)$\,, $(\vec{p}_\Lambda, \vec{p}_p)$\,, and 
$(\vec{p}_{\rho^0}\,,\vec{p}_{\pi^+})$\,, respectively, where 
$\vec{n}_{\Lambda_b}$ is the unit vector pointing toward the polarization of  $\Lambda_b$,
 $\vec{p}_{\Lambda, \rho^0}$ is the 3-momentum defined in the rest frame of $\Lambda_b$,  and $\vec{p}_p~(\vec{p}_{\pi^+})$   is defined in the helicity frame 
 of $\Lambda~(\rho^0)$. On the other hand, $\phi_1$~$(\phi_2)$ is  the azimuthal angle between the decay planes of $\hat{n}_{\Lambda_b}\times \vec{p}_\Lambda$ and $\vec{p}_\Lambda \times \vec{p}_p$~$(\vec{p}_{\rho^0} \times \vec{p}_{\pi^+})$.
The angular distributions are then given by
\begin{eqnarray}\label{ADE}
{\cal D}(\vec{\Omega}) &\equiv& \frac{1}{\Gamma_c}\frac{\partial^5\Gamma_c}{\partial \cos \theta \partial \cos \theta_1  \partial \cos \theta_2 \partial \phi_1 \partial\phi_2 } = \frac{1}{
|a_+|^2+|a_-|^2+|b_+|^2+|b_-|^2
}\frac{3}{8\pi^2}\times \nonumber\\
&& \sum_{\lambda_1'\,,M = \pm 1/2}\rho_{M,M}
\left|
\sum_{\lambda_1\,, \lambda_2 } 
H_{\lambda_1\,,\lambda_2 }A\,_{\lambda_1'}
d^{\frac{1}{2}}(\theta)^{M}\, _{\lambda_1 - \lambda_2}
d^{\frac{1}{2}}(\theta_1)^{\lambda_1}\,_{\lambda_1'}
d^{1}(\theta_2)^{\lambda_2}\,_{0}
e^{i(\lambda_1\phi_1 +\lambda_2 \phi_2)} 
\right|^2
\end{eqnarray}	
where $\Gamma_c$  is the decay width of $\Lambda_b \to \Lambda(\to p\pi^-) \rho^0(\to\pi^+\pi^-)$,
 $\rho_{M,M}$ are the density matrix elements of $\Lambda_b$ in the polarized direction, 
$\rho_{\pm 1/2,\pm 1/2} = (1\pm P_b)/2$, with $P_b$ the polarized fractions,
and $A_{\pm} $ are  given by 
\begin{equation}
 \left|A_{ +}
\right|^2 = \frac{1+\alpha}{2}\,,\,\,\,\ \left|A_{ -}
\right|^2 = \frac{1-\alpha}{2}\,,
\end{equation}
with $\alpha$ the up-down asymmetry parameter for $\Lambda\to p \pi^-$~\cite{BESIII:2018cnd}.
For the charge conjugate decays, one has the same formula with $\overline{\alpha} = - \alpha$ by assuming CP conserved in $\Lambda\to p \pi ^ - $.
From the formalism, it is obvious that the complex phases of $A_\pm$
 would not affect the results.
Here, we see the merit of the helicity formalism in which the angles are untangled in the amplitudes, {\it i.e.}, the helicities are independent of the $\Lambda_b$ polarization and each other. 
In the next section, we will find that it is not the case, when the states
are expanded with the triple vector products.

\begin{figure}[t]\label{FIG1}
	\caption{The angles given in Eq.~\eqref{ADE}, where $\hat{n}_{\Lambda_b}$ is the unit vector in  the $\Lambda_b$ polarized direction. }
	\centering
	\includegraphics[width=0.9\textwidth]{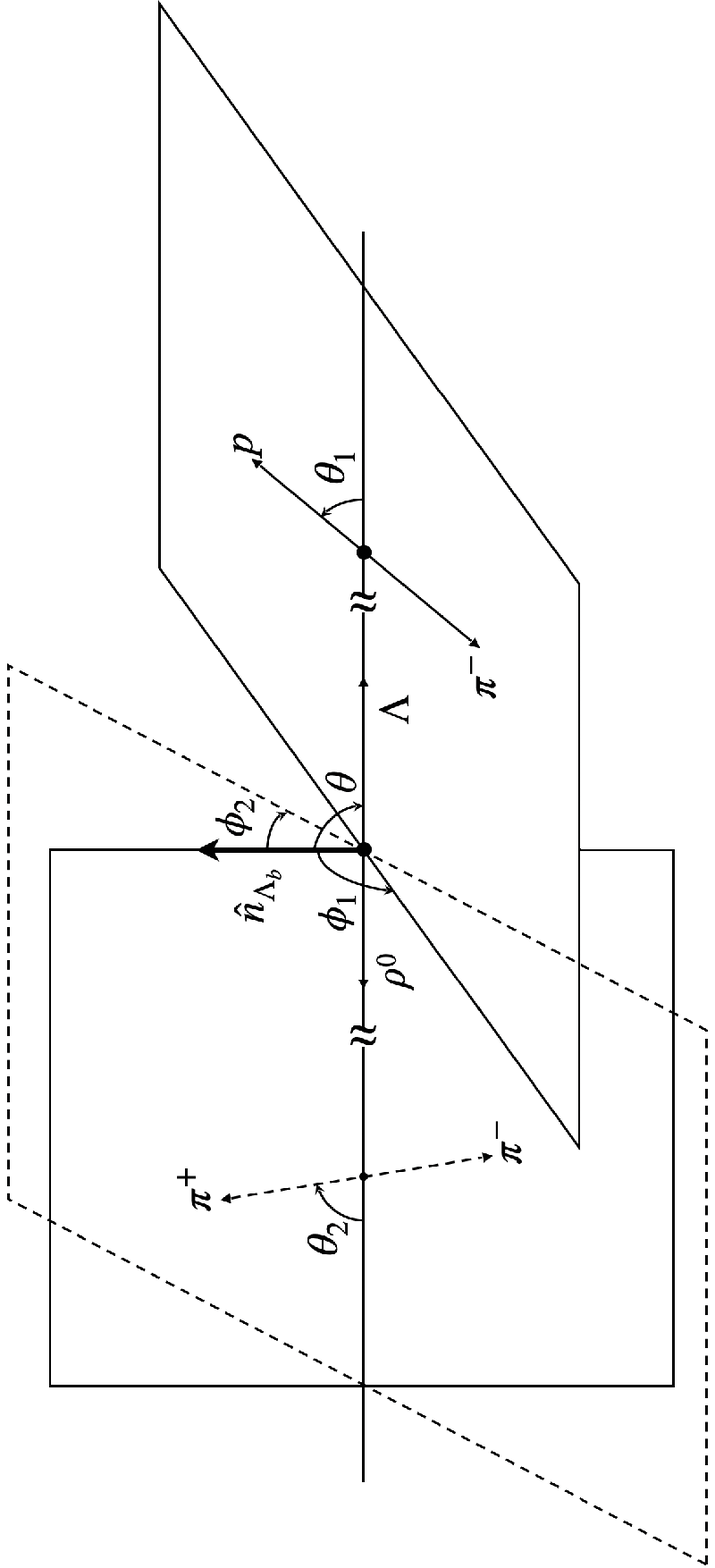}
\end{figure}

For the practical purpose,  ${\cal D}(\vec{\Omega})$  is  further written as
\begin{equation}\label{ADexpand}
{\cal D}(\vec{\Omega}) = \frac{1}{32\pi^2}
\frac{1}{|a_+|^2+|a_-|^2+|b_+|^2+|b_-|^2}
\sum^{20}_{i=1} f_i(a_\pm,b_\pm) D_i(\alpha_\Lambda,P_b, \theta\,, \theta_1\,, \phi_1 \,, \theta_2 \,, \phi_2 )\,, 
\end{equation}
where
$f_i$ are made of the amplitudes related to  $\Lambda_b \to \Lambda \rho^0$, and $D_i$ depend on the angles, $P_b$ and $\alpha_{\Lambda}$.
The explicit forms are given in Table~\MakeUppercase{\romannumeral 1}, where 
we have used the Legendre polynomial of $P_2=(3\cos ^2\theta_2  - 1 )/2 $ for the later convenience.
In general, ${\cal D}(\vec{\Omega})$ can
be applied to any sequential decay in the form of $1/2\to 1/2(\to 1/2, 0) 1 (\to0,0)$
with a straightforward replacement. For instance, the distribution of $\Xi_c^0 \to \Lambda(\to p \pi^-) \overline{K}^{*0}(\to \pi^+ K^-)$ can be given by substituting $\Xi_c^0$  and $\overline{K}^{*0}$ for $\Lambda_b$ and $\rho^0$, respectively~\cite{Belle:2021zsy}.

\begin{table}[t]\label{table1}
	\caption{
		The  parametrized angular distributions with the angles shown in FIG.1, where $P_2$ is  the Legendre polynomial with
		$P_2 = (3\cos^2 \theta_2 -1)/2$.}
	\begin{tabular}{l|cc}
		\hline
		$i$&$f_i$ & $D_i$ 	\\
		\hline
		1& $  |a_+|^2+|a_-|^2 + |b_+|^2 + |b_-|^2 $&$ 1 $ \\
		2&$ 2 |a_+|^2 +2|a_-|^2 - |b_+|^2 - |b_-|^2$&$P_2$\\
		3&$|a_+|^2- |a_-|^2-|b_+|^2 +|b_-|^2$&$\alpha \cos \theta_1$ \\
		4&$ 2 |a_+|^2 - 2|a_-|^2 + |b_+|^2 - |b_-|^2$&$\alpha \cos \theta_1 P_2$ \\
		5&$\frac{3}{\sqrt{2}}\Re(b_+a_+^*-b_-a_-^*)$&$\alpha \sin\theta_1 \sin(2\theta_2) \cos (\phi_1 + \phi_2) $\\
		6&$\frac{3}{\sqrt{2}}\Im(b_+a_+^*+b_-a_-^*)$&$\alpha \sin\theta_1 \sin(2\theta_2) \sin (\phi_1 + \phi_2) $\\
		7&$ |a_+|^2- |a_-|^2+|b_+|^2 -|b_-|^2$&$P_b \cos\theta $\\
		8&  $ 2 |a_+|^2 -2|a_-|^2 - |b_+|^2+ |b_-|^2$  &$P_b \cos \theta P_2$\\
		9&$\frac{3}{\sqrt{2}} 
		\Re ( a_+ b_-^{*} - a_-b_+ ^*)$
		&$P_b \sin \theta \sin(2\theta_2)\cos\phi_2$\\ 
		10&$\frac{3}{\sqrt{2}} 
		\Im ( a_+ b_-^{*} + a_- b_+^*)$
		&$P_b \sin \theta \sin(2\theta_2)\sin\phi_2$\\ 
		11&$ |a_+|^2+|a_-|^2 - |b_+|^2 - |b_-|^2$&$ P_b \alpha \cos \theta \cos \theta_1  $\\
		12& $ 2 |a_+|^2 +2|a_-|^2 + |b_+|^2+ |b_-|^2$  &$P_b\alpha  \cos \theta \cos \theta_1  P_2$\\
		
		13&$\frac{3}{\sqrt{2}}\Re (b_+a_+^* +b_- a_-^*)$&$P_b\alpha  \cos\theta \sin \theta_1 \sin(2 \theta_2)\cos(\phi_1 + \phi_2)$\\
		14&$\frac{3}{\sqrt{2}}\Im (b_+a_+^* - b_-a_-^*)$&$P_b\alpha  \cos\theta \sin \theta_1 \sin(2 \theta_2)\sin(\phi_1 + \phi_2)$\\
		15&$\frac{3}{\sqrt{2}}\Re (a_+b_-^* + a_-b_+^*)$&$P_b \alpha \sin\theta \cos \theta_1 \sin(2\theta_2) \cos \phi _2$\\
		16&$\frac{3}{\sqrt{2}}\Im (a_+b_-^* -a_- b_+^*)$&$P_b \alpha \sin\theta \cos \theta_1 \sin(2\theta_2) \sin \phi _2$\\
		17&$ -2 \Re ( a_- a_+^*)$&$P_b \alpha \sin \theta \sin \theta_1(1+2P_2) \cos \phi_1$\\
		18&$ -2 \Im ( a_- a_+^*)$&$P_b \alpha \sin \theta \sin \theta_1 (1+2P_2)\sin \phi_1$\\
		19&$2\Re(b_+b_-^*)$&$P_b \alpha  \sin \theta \sin \theta_1 (1- P_2) \cos (\phi_1 + 2 \phi_2 )$\\
		20&$2\Im(b_+b_-^*)$&$P_b \alpha  \sin \theta \sin \theta_1 (1- P_2) \sin  (\phi_1 + 2 \phi_2 )$\\
		\hline
	\end{tabular}
\end{table}
\label{AD}

In  ${\cal D}(\vec{\Omega})$,  $f_i$ depend on 4 complex amplitudes, in which a real parameter and  a complex phase could be diminished by the normalization, resulting in 3 real parameters and 3 relative complex phases left. Although the explicit forms are complicated, there have some rules which  ${\cal D}(\vec{\Omega})$  follows:
\begin{itemize}
\item  In the unpolarized case, $P_b = 0 $,  ${\cal D}(\vec{\Omega})$  shall only depend on $\theta_1$, $\theta_2$ and  $ \phi_1 + \phi_2$~(see FIG.~1). 
\item Since $\rho^0$ decays strongly,  ${\cal D}(\vec{\Omega})$  is invariant under interchanging $\pi^\pm$ in the cascade decay of $\rho^0$. Therefore, it does not matter which pseudoscalar meson  we choose to define $\theta_2$. Formally, it means that $D_i$ is invariant under the transformation of $(\phi_2\,,\theta_2)\to ( \phi_2 + \pi\,, \pi - \theta_2)$.
\item On the other hand, $\Lambda$ decays weakly, and hence it is important that $\theta_1$ is chosen as the polar angle between $\hat{n}_{\Lambda_b}$ and $\vec{p}_\Lambda$. Formally, $D_i$ is unaltered under the transformation of $(\phi_1,\theta_1,\alpha) \to (\phi_1+\pi\,,\pi - \theta_1\,, -\alpha) $.
\end{itemize}
Even when $\Lambda_b$  is unpolarized, it is still possible to determine $|a_\pm|^2$ and $|b_\pm|^2$ with $f_{2,3,4}$ and the decay widths. Furthermore, the relative phase between $b_+~(b_-)$ and $a_+~(a_-)$ can be extracted from $f_{5,6}$.

To reduce the uncertainties caused by $P_b$, one can sum over the degree of freedoms in the $\Lambda_b$ polarization.
To this end, one defines that
\begin{eqnarray}
\chi_1 &=& \phi_1 + \phi_2\,,\,\,\,\,\,\,\,\,\,0<\chi_1 <2\pi\,, \nonumber\\
\chi_2 &=& \frac{1}{2}(\phi_1 -\phi_2)\,,\,\,\,\,\,\,-\pi <\chi_2 <\pi\,.
\end{eqnarray}
The unpolarized angular distribution is then given as 
\begin{equation}\label{UNpolarizedAD}
{\cal D}^0 ( \vec{\Omega}^0) = \int^{\pi}_{-\pi} \int ^1_{-1}
{\cal D}^\chi ( \vec{\Omega}^\chi) 
d\cos \theta d \chi_2\,,
\end{equation}
where only $D_i$ with $i\le 6$ survive from the selection rule.
 In analogy to Eq.~(\ref{ADE}), we have 
\begin{eqnarray}\label{ADEun}
{\cal D}^0(\vec{\Omega}^0) &=& \frac{1}{\Gamma_c} \frac{\partial^3 \Gamma_c}{\partial \cos \theta_1 \partial \cos \theta_2 \partial \chi_1}\nonumber\\
&=&\frac{1}{8\pi}\frac{1}{|a_+|^2+|a_-|^2+|b_+|^2+|b_-|^2}\sum_{i=1}^6 f_i(a_\pm,b_\pm) D_i^0(\alpha_\Lambda\,, \theta_1\,, \theta_2 \,,\chi_1 )\,,
\end{eqnarray}
which is clearly independent of $P_b$.
Here, ${\cal D}^\chi( \vec{\Omega}^\chi) $ and $D_i^0$ are obtained by changing the parametrization of the azimuthal angles from $(\phi_1, \phi_2)$ to $(\chi_1,\chi_2)$ in ${\cal D}(\vec{\Omega})$ and $D_i$, respectively.

By integrating $ \theta_2$ and $\chi_1$ in ${\cal D}^0(\vec{\Omega}^0)$, we obtain that
\begin{eqnarray}\label{LambdaA}
\frac{1}{\Gamma_c} \frac{d\Gamma_c}{d\cos \theta_1}&=& (1 + \alpha_{\lambda_1} \alpha \cos \theta_1)/2 \,,
\end{eqnarray}
where
\begin{equation}\label{alpha1}
\alpha_{\lambda_1} \equiv \frac{\Gamma(\lambda_1 = 1/2) -\Gamma(\lambda_1 =- 1/2)   }{\Gamma(\lambda_1 = 1/2) +\Gamma(\lambda_1 = -1/2) }=\frac{|a_+|^2+|b_-|^2- |a_-|^2-|b_+|^2 }{|a_+|^2+ |a_-|^2+|b_+|^2 +|b_-|^2}\,,
\end{equation}
describing the polarization  asymmetry of $\Lambda$. 
Likewise, we can integrate $\theta_1$ and $\chi_1$, given by
\begin{equation}\label{LambdaA2}
\frac{1}{\Gamma} \frac{d\Gamma}{d\cos \theta_2}= \frac{1}{2} + \frac{P_2}{4} + \frac{3}{4}P_2\alpha_{\lambda_2}  \,,
\end{equation}
where
\begin{equation}\label{alpha2}
\alpha_{\lambda_2} \equiv \frac{ |a_+|^2 +|a_-|^2 - |b_+|^2 - |b_-|^2
}{|a_+|^2+ |a_-|^2+|b_+|^2 +|b_-|^2}\,,
\end{equation}
representing the asymmetry between the longitudinal and transverse polarizations of $\rho^0$.
With Eq.~\eqref{Is}, it is straightforward to see that $\alpha_{\lambda_1}$ and $\alpha_{\lambda_2}$ are P-odd and P-even, respectively.
In addition, $\alpha_{\lambda_{1,2}}$ are independent of $P_b$, which is a reasonable result since $\lambda_{1,2}$ are $SO(3)_R$ scalars and shall not be affected by the spin direction of $\Lambda_b$ alone.

The structures of the angular distributions can be traced back to the spin correlations in $\Lambda_b \to \Lambda V$.
 In the next section, we will study the T-odd spin correlations and identify their effects on  ${\cal D}(\vec{\Omega})$ .

In the near future, it is not likely that the experiments have enough data points to reconstruct the full angular distributions in Eq.~\eqref{ADE}. Nonetheless, the experiments at LHCb are able to carry out the partial angular distribution
analyses, which require a handful of parameters to be fitted, such as the ones in Eqs.~\eqref{ADEun}, \eqref{LambdaA} and \eqref{LambdaA2} (see Eq.~\eqref{recommand} also). In particular, aiming on probing T-violation, the partial angular analysis of $\Lambda_b\to \Lambda \phi$ has been studied at LHCb.~\cite{LambdabLambdaPhiExp}, although the effects of the FSIs have not been considered.

\section{T-odd observables}
In general, to observe  T-violating effects, one should compare $\Lambda_b \to \Lambda V$ to the time-reversed processes, $\Lambda V \to \Lambda_b$,  which are difficult to prepare in the experiments. However, in the first order of the weak interaction, it is possible to constrain the amplitudes with the T symmetry as demonstrated in the helicity formalism~(see Eq.~\eqref{usualTR} and Appendix A also). 
Our goal in this section is to obtain naive T-odd observables with Eq.~\eqref{appenreal} and subtract the effects of the FSIs by comparing them with the charge conjugate ones. 
To this end, we ought to find
the eigenstates of the T-odd operators, $\hat{T}_i$,  in the $\Lambda V$ systems,  which satisfy
\begin{equation}\label{criteria}
I_t |\Lambda V; \lambda_{t_i}, \{\lambda_f\} \rangle = |\Lambda V; -\lambda_{t_i}, \{\lambda_f^T\} \rangle \,,
\end{equation}
where $I_t$ is the T operator, $\lambda_{t_i}$ are the eigenvalues of $\hat{T}_i$,  $\{\lambda_f\}$ corresponds to  a set of compatible physical quantities to specify the states, and $\lambda_f^T$ represent the values after  the T transformation.

From Eq.~\eqref{A2},
if ${\cal H}_{eff}$ respects the T symmetry, we have
\begin{equation}\label{newadd}
|\langle \Lambda V; \lambda_{t_i} , {\text{``out''}} , \{ \lambda_f\} | {\cal H}_{eff} | B\rangle |^2 -  |\langle \Lambda V; -\lambda_{t_i} , {\text{``in''}} , \{ \lambda_f^T\} | {\cal H}_{eff} | B\rangle |^2 =0  \,.
\end{equation}
In our study,
except for $J_z$ which can always be flipped back by $SO(3)_R$\footnote{See the footnote followed by Eq.~\eqref{usualTR}},
$\lambda_f$ are chosen as T-even.
The naive T-violating observables are then given by
\begin{equation}\label{naive}
\Delta_{t_i} = \frac{\Gamma(\lambda_{t_i})-\Gamma(-\lambda_{t_i})}{\Gamma(\lambda_{t_i})+\Gamma(-\lambda_{t_i})}\,,
\end{equation}
where we always choose $\lambda_{t_i}>0$  to fix the ambiguity. Notice that if ``in'' and ``out'' can be interchanged freely, 
the left-handed side of Eq.~\eqref{newadd} would be proportional to the the numerator of $\Delta_{t_i}$, resulting in  that $\Delta_{t_i}$  are T-odd observables. Furthermore,
since $\hat{T}_i$  involve more than two spins in two-body decays, $\Delta_{t_i}$ are also  referred to as the T-odd spin correlations.

Nonetheless,  $|\pm\lambda_{t_i}\rangle$ could potentially oscillate, leading to
\begin{equation}
\langle -\lambda_{t_i} ; {\text{``out''}}| \lambda_{t_i} ; {\text{``in''}}\rangle \neq 0 \,,
\end{equation}
for the rescattering from $\lambda_{t_i} \to -\lambda_{t_i}$, due to the FSIs. To subtract the rescattering effects, 
we have to compare them with their charge conjugates, as long as the FSIs are not tractable. Accordingly, the 
true T-violating quantities can be given as 
\begin{equation}
{\cal T}_{t_i}  \equiv (\Delta_{t_i} - (\pm)\overline{\Delta}_{t_i})/2\,,
\end{equation}
where the overline denotes the charge conjugate, and
the signs correspond to the parities of $\hat{T}_i$.
 For example, if  $\hat{T}_j$ are the vector products of two spins and one 3-momentum, it is P-odd with the minus sign for ${\cal T}_{t_j}$.
Here, as we have used the CPT symmetry to relate the charge conjugate processes,  ${\cal T}_{t_i}$ are not only T-violating  but also CP-violating
observables.

To define the T-odd operators, we start with the definition of the spin, given by~\cite{spinoperators}
\begin{equation}\label{spin}
m  \vec{s} =P^0 \vec{J} - \vec{p}\times \vec{K} - \frac{1}{P^0+m} \vec{p}(\vec{p}\cdot\vec{J})\,,
\end{equation}
where $m$ is the particle's mass, and $\vec{K}$ is the generator of the Lorentz boost.
Despite that the definition seems complicated, 
each term can be understood separately. 
On the right-hand side of Eq.~\eqref{spin},
the first term describes that $\vec{s}$ shall be reduced to $\vec{J}$ at the rest frame,  the second one ensures $[s_i,p_j]=0\,,$  and the last one guarantees that the algebra is closed, $i.e.$, $[s_i,s_j]= i\epsilon_{ijk}s_k$. As a result, we have~\cite{spinoperators}
\begin{equation}\label{spinoperator}
s_i L|\vec{p}=0, J_z = M\rangle = L J_i |\vec{p}=0, J_z = M\rangle \,,
\end{equation}
where $L$ is an arbitrary Lorentz boost,
 indicating that $\vec{s}$ is treated as a 3-momentum state of $\vec{J}$ in its rest frame.

In quantum mechanics, one should find a set of commuting operators to characterize the states properly, which has often been mishandled in  the studies concerning T-violation. For instance, let us consider the most simple T-odd operator, given by
\begin{equation}\label{T1}
\hat{T}_1 = (\vec{s}_1 \times \vec{s}_2) \cdot \hat{p}_1\,,
\end{equation}
where $\vec{s}_1$ and $\vec{s}_2$ are the spins of $\Lambda$ and $V$, respectively, and $\hat{p}_1$ is the normalized unit vector of $\vec{p}_1$. In the literature, it is often examined in terms of the Dirac spinors and polarization vector as Eq.~\eqref{badEXP}, which essentially expand the final states by $(\vec{p}_1,\vec{s}_1\cdot \hat{n}_1)$ and $(\vec{p}_2,\vec{s}_2\cdot \hat{n}_2)$\,, where $\hat{n}_1$ and $\hat{n}_2$ are arbitrary unit vectors in the three-dimensional space.
However, the approach is questionable since $\vec{s}_1 \cdot \hat{n}_1$ does not commute with $\hat{T}_1$. It is odd~(wrong) to study any physical quantity with incompatible observables.

For each T-odd operator, one should find a set of compatible observables, in which $J^2$ and $J_z$ are always included, resulting in that only T-odd $SO(3)_R$ scalars are studied.  These T-odd scalars are classified into 4 categories, depending on the spins. First, we study 
$\hat{T}_1$ defined in Eq.~\eqref{T1}\,,
which can be sizable even when $P_b = 0 $, since it does not contain $\vec{J}$. Second, we work out the case in which $\vec{s}_1$ is not involved, given as 
\begin{equation}\label{T2}
\hat{T}_2 = (\vec{s}_2 \times \hat{p}_1 ) \cdot \vec{J}\,,
\end{equation}
which can be large even when $\alpha = 0$.
Third, 
we discuss the case with $\vec{s}_1$, given as
\begin{equation}\label{T3}
\hat{T}_3 \equiv  (\vec{s}_1 \times \hat{p}_1 ) \cdot \vec{J}\,,
\end{equation}
which requires both $P_b \neq 0 $ and $\alpha \neq  0$. Note that if we set $\alpha = 0$, $D_i$ will be independent of $\theta_1$ and $\phi_1$ as in the cascade decays of $1/2\to 1/2, 0$  shown in Eq.~\eqref{LambdaA}.
Finally, we examine the cases in which all the spins are involved.

\subsection{T-odd observables with $\vec{s}_1$ and $\vec{s}_2$}
In this subsection, we consider the T-odd scalar operators made from $\vec{s}_1\,,\vec{s}_2$, $\hat{p}_1$ and $\vec{J}\cdot \hat{p}_1$.  Since $\hat{p}_1$ commutes with each of them,  the T-odd scalar operators must also commute with $\vec{J}\cdot\hat{p}_1$. The most simple case is $\hat{T}_1$, given in Eq.~(\ref{T1}).
 By direct computation with Eqs.~(\ref{HelicityStates}) and (\ref{spinoperator}), the eigenstates are given by
\begin{eqnarray}\label{lambdat1}
|\lambda_{t_1}= \pm \frac{1}{\sqrt{2}}\,,\lambda=\frac{1}{2} \rangle =  \frac{1}{\sqrt{2}}\left(
| a_+\rangle  \mp i 
|b_+\rangle 
\right)\,,\nonumber\\
|\lambda_{t_1}= \pm \frac{1}{\sqrt{2}}\,,\lambda=-\frac{1}{2} \rangle =  \frac{1}{\sqrt{2}}\left(|a_-\rangle \pm i
| b_-\rangle 
\right)\,,
\end{eqnarray}
where $\lambda_{t_1}$ and $\lambda = \lambda_1 -\lambda_2$ are the eigenvalues of $\hat{T}_1$ and  $\vec{J}\cdot \hat{p}_1$, respectively. 
Due to the anti-linear property of $I_t$, one can easily check that
the eigenstates in Eq.~\eqref{lambdat1} satisfy the criterion in Eq.~(\ref{criteria}).

In addition, it is straightforward to see the helicities are entangled in $|\lambda_{t_1}\rangle$. Nonetheless, they are still independent of $\hat{n}_{\Lambda_b}$. 
Since $\hat{T}_1$ commutes with $\vec{J}\cdot\hat{p}_1$, $|a_+\rangle$ and $|a_- \rangle$ share the same  $\theta$ dependence with  $|b_+\rangle$ and $|b_- \rangle$, respectively. For instance, from Eq.~\eqref{HelicityStates}, we have
\begin{equation}\label{eigenstatest1}
|\lambda_{t_1}= \pm \frac{1}{\sqrt{2}}\,,\lambda=\frac{1}{2} \rangle = 
\frac{1}{2\sqrt{2 }\pi}
\int d\Omega 
\left(
 |\vec{p}_1, 1/2\,, 0\rangle \mp i  |\vec{p}_1, -1/2\,,\mp 1\rangle   
\right)
 e^{-iM\phi} d^\frac{1}{2}(\theta) ^{M}\,_{\frac{1}{2}}\,.
\end{equation}
Clearly, the helicities do not depend on $\theta$ and hence $\hat{n}_{\Lambda_b}$.

Notice the close relations between the opposite $\lambda$. The eigenstates in Eq.~\eqref{lambdat1} are related by the parity, which can be seen from the following identity, read as~\footnote{
In practice, $J_z$ flips the sign under $I_t$. However, we can always rotate it back with $R_2(-\pi)$ without affecting $\lambda_{t_i}$ and $\lambda$, and therefore, the conclusion still holds.
}
 \begin{equation}\label{argument}
I_t I_s |\lambda_{t_1} , \lambda\rangle =I_{t} | -\lambda_{t_1}, -\lambda\rangle = |\lambda_{t_1}, -\lambda\rangle \,.
	\end{equation}
Here, the first and second
 equalities  are due to that $\lambda_{(t_1)}$ and $\lambda$ are P-odd and T-even, respectively.
In general, $\lambda_{t_1}$ is  degenerated as long as $\lambda \neq 0$.

The naive T-odd observable, defined in Eq.~\eqref{naive}, is then given as
\begin{eqnarray}\label{naivet1}
\Delta_{t_1} \equiv  \frac{\Gamma(\lambda_{t_1})-\Gamma(-\lambda_{t_1})}{\Gamma(\lambda_{t_1})+\Gamma(-\lambda_{t_1})} = \frac{2\Im (a_+b_+^*-a_-b_-^*)}{|a_+|^2+|a_-|^2+|b_+|^2+|b_-|^2}\,,
\end{eqnarray}
which is also P-odd. Consequently, the true T-odd quantity is given by
\begin{equation}\label{TrueT1}
{\cal T}_{t_1} = (\Delta_{t_1} + \overline{\Delta}_{t_1})/2\,.
\end{equation}
From Eq.~(\ref{naivet1}), it is obvious that $\Delta_{t_1}$ vanishes if $a_\pm$ and $b_\pm$ are real. This is consistent with Eq.~\eqref{usualTR}.

To construct a P-even and T-odd observable, we define
\begin{equation}\label{naivet1p}
  \hat{T_1}^p=\vec{J}\cdot\hat{p}_1\hat{T_1}\,.
  \end{equation}
By comparing Eqs.~\eqref{naivet1p} and \eqref{eigenstatest1}, it is easy to see that
  the eigenstates of $\hat{T}^p_1$ are identical to those of $\hat{T}_1$ with the eigenvalues of $\lambda_{t_1}^p = \lambda_{t_1}\lambda$. 
Accordingly, the naive T-odd observable is then  given as 
\begin{eqnarray}
\Delta_{t_1}^{p} =
\frac{2\Im (a_+b_+^*+a_-b_-^*)}{|a_+|^2+|a_-|^2+|b_+|^2+|b_-|^2}\,,
\end{eqnarray}
and the true one is 
\begin{equation}\label{TrueT1p}
{\cal T}_{t_1}^p = (\Delta^p_{t_1} - \overline{\Delta}^p_{t_1})/2\,.
\end{equation}

With the unpolarized $\Lambda_b$, there are only 3 independent observables in the decay distributions, $\vec{p}_a\,, \vec{p}_b$ and $\vec{p}_c$, where $a\,,b\,$ and $c$ correspond to the particles after the cascade decays.  As a result, to manifest the T-odd quantity, it can only take the P-odd form, $(\vec{p}_a\times \vec{p}_b)\cdot \vec{p}_c$. 
Hence, one naively expects that  $\Delta_{t_1}$~(P-odd) shows up  in ${\cal D}^0(\vec{\Omega}^0)$, whereas $\Delta_{t_1}^p$~(P-even) can not be observed. It is indeed the case in our previous work~\cite{Geng:BtoVV}, in which we consider $B\to VV'$.
Interestingly, things go the other way round in $\Lambda_b$. 
We see that $\Delta_{t_1}$ appears in $f_{14}$, which requires $P_b\neq 0$, whereas
$\Delta_{t_1}^p$ is found in $f_6$, which is independent of $P_b$.
The reason for this opposite behavior is that to manifest the helicities of $\Lambda$, $\alpha$ is always needed~(see Eq.~(\ref{LambdaA})), which is P-odd, and therefore inverts the argument.

Since the observation of  $\Delta_{t_1}^p$  does not demand $\Lambda_b$ to be polarized, the uncertainties  caused by $P_b$ can be eliminated.
From ${\cal D}^0(\vec{\Omega}^0)$, we find that
\begin{equation}\label{recommand}
\Delta^p_{t_1}= -\frac{128\sqrt{2}}{3\pi^2\alpha}
\int^{2\pi}_0\int^1_{-1}\int^1_{-1} {\cal D}^0 ( \vec{\Omega}^0) d\cos \theta_1 \left(\cos\theta_2 d\cos \theta_2 \right)
\left(
\sin \chi_1 d\chi_1
\right)  \,.
\end{equation}
Note that in contrast to the ordinary integral, we have added the weight factors, $\cos \theta_2$ and $\sin\chi_1$. Here, $\cos\theta_2$ is designed to diminish the uncertainties, satisfying the orthogonal relation, given as 
\begin{equation}
\int^1_{-1} D_i \cos \theta_2 d \cos \theta_2 = 0 \,\,\,\,\,\,\,\text{for}~~i = 1,2,3,4\,,
\end{equation}
while $\sin \chi_1$ corresponds to the familiar Fourier transformation.

\subsection{T-odd observables with $\vec{J}$ and $\vec{s}_2$}
Similar to the previous subsection, we consider the T-odd scalar operators made of $\vec{s}_2$, $\hat{p}_1$,  $\vec{J}$ and $\vec{s}_1 \cdot \hat{p}_1$. Since $\vec{s}_1 \cdot \hat{p}_1$ commutes with each of them, the T-odd operators  commute with $\vec{s}_1 \cdot \hat{p}_1$. 
From Eq.~\eqref{argument} with substituting $\lambda_1$ for $\lambda$,  we anticipate that the eigenvalues are degenerated as $\lambda_1\neq 0 $.
For the most simple case of $\hat{T}_2$ in Eq.~\eqref{T2}, we have that
\begin{eqnarray}
|\lambda_{t_2}= \pm 1/\sqrt{2}\,,\lambda_1=1/2 \rangle =  \frac{1}{\sqrt{2}}\left(
|a_+\rangle \pm i | b_-\rangle 
\right)\,,\nonumber\\
|\lambda_{t_2}= \pm 1/\sqrt{2}\,,\lambda_1=-1/2 \rangle =  \frac{1}{\sqrt{2}}\left(
 | a_-\rangle \mp  i |b_+\rangle  
\right)\,,
\end{eqnarray}
where $\lambda_{t_2}$ and $\lambda_{1}$ are the eigenvalues of 
$\hat{T}_2$ and $\vec{s}_1\cdot \hat{p}_1$, respectively, while the naive T-odd observable is
\begin{eqnarray}
\Delta_{t_2} =\frac{ 2\Im (b_-a_+^*-b_+a_-^*)}{|a_+|^2+|a_-|^2+|b_+|^2+|b_-|^2}\,,
\end{eqnarray}
where the spins correlations are handed down to $f_{10}$. 
With the charge conjugate,
the true T-odd observable is then given as 
\begin{equation}\label{TrueT2}
{\cal T}_{t_2} =
(\Delta_{t_2} + \overline{\Delta}_{t_2} )/2\,.
\end{equation}

In analogy to the previous subsection, we define $\hat{T}^p_2= \vec{s}_1\cdot \hat{p}_1 \hat{T}_2 $, resulting in the naive T-odd observable as 
\begin{eqnarray}
\Delta_{t_2}^P =\frac{ 2\Im (b_-a_+^*+b_+a_-^*)}{|a_+|^2+|a_-|^2+|b_+|^2+|b_-|^2}\,,
\end{eqnarray}
which can be found in $f_{16}$, and the true T-odd observable is 
\begin{equation}\label{TrueT2p}
{\cal T}_{t_2}^p = (\Delta_{t_2}^p -  \overline{\Delta}_{t_2}^p)/2\,.
\end{equation}
In Sec.~\MakeUppercase{\romannumeral 4}, we will see that $\Lambda_b \to \Lambda V$ is predominant by $a_-$ and $b_+$, making $\Delta_{t_2}^{(p)}$ a good observable to test the SM.

\subsection{T-odd observables with $\vec{J}$ and $\vec{s}_1$}
We now expand the states in terms of $\vec{s}_2\cdot \hat{p}_1$ and $\hat{T}_3$.
 The eigenstates with nonzero eigenvalues are 
\begin{equation}
|\lambda_{t_3}= \pm \frac{1}{2}\rangle = \frac{1}{\sqrt{2}}\left(
|a_+\rangle \pm i | a_-\rangle 
\right)\,,
\end{equation}
where $\lambda_{t_3}$ are the eigenvalues of $\hat{T}_3$.
On the other hand, similar to Eq.~\eqref{argument}, we have 
\begin{equation}
\hat{T}_3 |b_\pm \rangle = 0\,.
\end{equation}
The naive T- odd observable is 
\begin{equation}
\Delta_{t_3} \equiv \frac{2\Im(a_-a_+^*)}{|a_+|^2+|a_-|^2+|b_+|^2+|b_-|^2}\,,
\end{equation}
which manifests itself in $f_{18}$. The true T-odd observable is 
\begin{equation}\label{TrueT3}
{\cal T}_{t_3} =( \Delta_{t_3} - \overline{\Delta}_{t_3})/2\,,
\end{equation}
which is also a CP-violating observable due to the CPT symmetry.

\subsection{T-odd observables with triple spin correlations}
Among the T-odd parameters in $D(\vec{\Omega})$,
only $f_{20}$ has not been discussed yet.
The responsible T-odd operator is quite complicated, read as
\begin{equation}
\hat{T}_4=
\left[
\hat{T}_1 \left(
\vec{s}_2 \cdot \vec{J} -\frac{1}{2}(\vec{s}_2 \cdot\hat{p}_1)^2
\right)
+\frac{1}{\sqrt{2}} \hat{T}_3 
\right] + h.c.\,,
\end{equation}
where $h.c.$ stands for the Hermitian conjugate. The eigenstates are
\begin{equation}
 | \lambda_{t_4} =\pm \sqrt{2}\rangle  = \frac{1}{\sqrt{2}}\left(
 |b_+\rangle \pm i |b_-\rangle
 \right)\,,
\end{equation}
where $\hat{T}_4|a_\pm\rangle =0 $.
The corresponding naive T-odd observable is then given as
\begin{equation}
\Delta_{t_4} \equiv \frac{2\Im(b_-b_+^*)}{|a_+|^2+|a_-|^2+|b_+|^2+|b_-|^2}\,,
\end{equation}
which is proportional to the relative phase between $b_\pm$,
while the true T-odd observable is
\begin{equation}\label{TrueT4}
{\cal T}_{t_4} = (\Delta_{t_4} + \overline{\Delta_{t_4}})/2\,.
\end{equation}
Here, the nonzero value of $\Delta_{t_4}$ could be caused by the FSIs. However, to oscillate between $|b_\pm\rangle$, the helicities of V must alter twice~$(\lambda_2 : 1\to 0 \to -1)$, and therefore, the oscillations are expected to be suppressed, as the case in $B$ mesons decays~\cite{Geng:BtoVV}. 
Clearly, it is interesting to see whether the suppression holds in the baryon systems or not.

Before ending this section, let us collect the results. We have found the T-odd spin correlations in $\Lambda_b \to \Lambda V$, with their effects on the sequential decays identified.
In particular, we have shown that $\Delta_{t_1}\,, \Delta_{t_1}^p\,,\Delta_{t_2}\,,\Delta_{t_2}^p\,,\Delta_{t_3}$ and $\Delta_{t_4}$ correspond to $f_{14}\,,f_{6}\,,f_{10}\,,f_{16}\,,f_{18}\,$ 
and $f_{20}$, respectively.
Notably, all the relative phases among $a_\pm$ and $b_\pm$ can be described by $\Delta_{t_{1,2,3,4}}^{(p)}$, which complete our study on  all possible T-odd observables.

\section{Numerical results}
In the SM,
the effective Hamiltonian responsible for $b\to d/s$ transitions, obtained from the operator product expansions, is given by~\cite{Buras:1991jm}
\begin{equation}
{\cal H}_{eff} = \frac{G_F}{\sqrt{2}}
\sum_{f=,d,s}\left[
V_{ub}V_{uf}^* \left(
C_1O^f_1 + C_2 O^f_2
\right) - V_{tb}V_{tq}^* \sum_{i=3}^{10}C_i O^f_i
\right]\,,
\end{equation}
where $G_F$ stands for the Fermi constant, $C_i$ represent the Wilson coefficients, $V_{qq'}$ correspond to the CKM matrix elements, and $O_{1-10}^f$ are the operator products, given by~\cite{Buras:1991jm}
\begin{eqnarray}
&&O_1^f =  (\overline{u}_\alpha b_\alpha)_L (\overline{f}_\beta u_\beta)_L \,,\,\,\,\,~~~~~~~~~~O_2^f =(\overline{u}_\alpha b_\beta)_L (\overline{f}_\beta u_\alpha)_L\,,\nonumber\\
&&O_3^f = (\overline{f}_\alpha b_\alpha)_L \sum_{q} (\overline{q}_\beta q_\beta)_L\,,\,\,\,\,~~~~~O_4^f = (\overline{f}_\alpha b_\beta)_L \sum_{q}(\overline{q}_\beta q_\alpha)_L\,,\nonumber\\
&&O_5^f = (\overline{f}_\alpha b_\alpha)_L \sum_{q}(\overline{q}_\beta q_\beta)_R\,,\,\,\,\,~~~~~O_6^f = (\overline{f}_\alpha b_\beta)_L \sum_{q}(\overline{q}_\beta q_\alpha)_R\,,\nonumber\\
&&O_7^f = \frac{3}{2}(\overline{f}_\alpha b_\alpha)_L \sum_{q}e_{q}(\overline{q}_\beta q_\beta)_R\,,\,\,\,\,O_8^f =\frac{3}{2} (\overline{f}_\alpha b_\beta)_L \sum_{q}e_{q}(\overline{q}_\beta q_\alpha)_R\,,\nonumber\\
&&O_9^f = \frac{3}{2}(\overline{f}_\alpha b_\alpha)_L \sum_{q}e_{q} (\overline{q}_\beta q_\beta)_L\,,\,\,\,\,O_{10}^f =\frac{3}{2} (\overline{f}_\alpha b_\beta)_L \sum_{q}e_{q}(\overline{q}_\beta q_\alpha)_L\,,
\end{eqnarray}
with 
$L$ and $R$ in the subscripts denoting the left and right-handed currents, respectively. The amplitudes are  
given by sandwiching ${\cal H}_{eff}$ between the initial and final states.
Note that both $C_i$ and $O_i$ depend on the renormalization schemes and energy scales.

In the following, we adopt the generalized factorization approach~\cite{EffectiveWilson}, in which the quarks and antiquarks of vector mesons are created 
by weak vertices. The amplitudes are simplified as~\cite{KornerSM,Zhu:2018jet}
\begin{equation}\label{Factorized}
\frac{G_F}{\sqrt{2}}{\cal C}_V \langle V| V^\mu|0\rangle \langle\Lambda|(\overline{s}b)_L| \Lambda_b\rangle=
\frac{G_F}{\sqrt{2}}{\cal C}_V f_V M_V\xi^{\mu *}\langle\Lambda|(\overline{s}b)_L\gamma_\mu b| \Lambda_b\rangle\,, 
\end{equation}
where $V^\mu$,
$f_V$, $M_V$, and $\xi^\mu$ are the  currents, decay constants, masses, and polarizations of the vector mesons $(V)$, and ${\cal C}_V$ are given as
\begin{eqnarray}
{\cal C}_{\phi} &=& -V_{ts}^*V_{tb} \left(
a_3+a_4 +a_5 -\frac{1}{2}a_7-\frac{1}{2}a_9-\frac{1}{2}a_{10}
\right)\,,\nonumber\\
{\cal C}_{\rho^0} &= &
\frac{1}{\sqrt{2}}\left[
V_{us}^*V_{ub}a_2-
\frac{3}{2}V_{ts}^*V_{tb} 
\left(
a_7+
a_9 
\right)
\right]\,,\nonumber\\
{\cal C}_{\omega} &= &
\frac{1}{\sqrt{2}}\left\{
V_{us}^*V_{ub}a_2
-
 V_{ts}^*V_{tb} 
\left[
2a_3+2a_5
+
\frac{1}{2} 
\left(
a_7+
a_9  
\right)
\right]
\right\} \,,\nonumber\\
{\cal C}_{K^{*0}} &= &
-
V_{td}^*V_{tb} 
\left(
a_4-\frac{1}{2}a_{10} 
\right)\,,
\end{eqnarray}
respectively, with $a_i = c_i^{eff} + c_{i+(-1)}^{eff}/N_c$ for $i=$odd~(even). Here, one has $N_c=3$ in the absence of the NF contributions. In the numerical calculations, the Wolfenstein parametrization is used for the CKM matrix elements in the SM, taken to be~\cite{pdg}
\begin{equation}
\lambda=0.22650\pm 0.00048\,,\,\,\,A=0.790^{+0.017}_{-0.012}\,,\,\,\,\rho = 0.141^{+0.016}_{-0.017}\,,\,\,\,\eta = 0.357\pm 0.011\,,
\end{equation}
and the effective Wilson coefficients are given in Table~\MakeUppercase{\romannumeral 2}~\cite{GeneralGeng:2021nkl,EffectiveWilson}.

\begin{table}\label{EffectiveWilsonCoefficientTable}
	\caption{The effective Wilson coefficients with the NDR and $\overline{\text{MS}}$ schemes  at the energy scale of $\mu = 2.5$\,GeV, where 
		$c_{1,2}^{eff}$ and $c_{3-10}^{eff}$ are in the units of $10^0$ and $10^{-4}$, respectively.}
	\begin{tabular*}{0.8\textwidth}{c @{\extracolsep{\fill}} cc|cc}
		\hline
		&$b\to d$&$\overline{b}\to \overline{d}	$	&$b\to s$&$\overline{b}\to \overline{s}$\\
		\hline
		$c_1^{eff}$&$1.168$&$1.168$&$1.168$&$1.168$\\
		$c_2^{eff}$&$-0.365$&$-0.365$&$-0.365$&$-0.365$\\
		$c_3^{eff}$&$238+14i$&$254+43i$&$243+31i$&$241+32i$\\
		$c_4^{eff}$&$-497-42i $&$-545-130i $&$-512-94i $&$-506-97i $\\
		$c_5^{eff}$&$145+14i$&$162+43i$&$150+31i$&$148+32i$\\
		$c_6^{eff}$&$-633-42i$&$-682-130i$&$-649-94i$&$-643-97i$\\
		$c_7^{eff}$&$-1.0-1.0i$&$-1.4-1.8i$&$-1.1-2.2i$&$-1.1-1.3i$\\
		$c_8^{eff}$&$5.0$&$5.0$&$5.0$&$5.0$\\
		$c_9^{eff}$&$-112-1i$&$-112-3i$&$-112-2i$&$-112-2i$\\
		$c_{10}^{eff}$&$20$&$20$&$20$&$20$\\
		\hline
	\end{tabular*}
\end{table}

The baryon transitions in Eq.~\eqref{Factorized}  can be parametrized by the form factors, defined by
\begin{eqnarray}
\langle \Lambda | \overline{s} \gamma^\mu b | \Lambda_b\rangle 
&=&\overline{u}_{\Lambda} \left(
f_1(q^2) \gamma^\mu - f_2 (q^2)i \sigma_{\mu \nu} \frac{q^\nu}{ M_{\Lambda_b}}   +f_3(q^2) \frac{q^\mu}{M_{\Lambda_b}}
\right)u_{\Lambda_b}\,,\nonumber\\
\langle \Lambda | \overline{s} \gamma^\mu \gamma^5 b | \Lambda_b\rangle 
&=&\overline{u}_{\Lambda} \left(
g_1(q^2) \gamma^\mu - g_2 (q^2)i \sigma_{\mu \nu} \frac{q^\nu}{ M_{\Lambda_b}}   +g_3(q^2) \frac{q^\mu}{M_{\Lambda_b}}
\right)\gamma^5 u_{\Lambda_b}\,,
\end{eqnarray}
where $q^\mu = p_{\Lambda_b}^\mu-p_{\Lambda}^\mu$, and $f_{1,2,3}$ and $g_{1,2,3}$ are the form factors. In this study, we calculate them  with the modified MIT bag model~\cite{GeneralModifiedBagModel}, in which the center motions of the baryon waves are removed. 
The hadrons parameters are given as~\cite{GeneralModifiedBagModel,Perez-Marcial:1989sch}
\begin{equation}
m_b = 4.8~\text{GeV},~~~~m_s = 0.28~\text{GeV},~~~~m_u=m_d=0.005~\text{GeV}\,,~~~~R^{-1}= 0.21\pm 0.01 \text{GeV}\,,
\end{equation}
where $R$ is the bag radius, and 
$(f_\phi,f_\omega,f_\rho) = (215\pm 5,187\pm5,216\pm 3)$~MeV~\cite{Ball:2004rg}\,.
The numerical results of the form factors with the $q^2$ dependencies are given in Table~\MakeUppercase{\romannumeral 3}. The uncertainties come mainly from the bag radius. We see that our results are consistent with those having $f_1=g_1$ demanded by the heavy quark symmetry. 

\begin{table}
	\caption{Form factors with the unit $10^{-1}$, where the uncertainties come from the bag radius.}
	\begin{tabular}{l| cccccc}
		\hline
		&$f_1$&$f_2$&$f_3$&$g_1$&$g_2$&$g_3$\\
		\hline
		$q^2 = m_\rho^2$&$1.60\pm 0.09$&$0.02\pm 0.00$&$-0.18\pm 0.00$&$1.60\pm 0.09$&$0.05\pm 0.00$&$-0.21\pm 0.00$\\
		$q^2 = m_{K^*}^2$&$1.62\pm 0.09$&$0.02\pm 0.00$&$-0.19\pm 0.01$&$1.62\pm 0.09$&$0.05\pm 0.00$&$-0.22\pm 0.01$\\
		$q^2 = m_\phi^2$&$1.63\pm 0.07$&$0.04\pm 0.01$&$-0.19\pm 0.01$&$1.64\pm 0.08$&$0.05\pm 0.01$&$-0.22\pm 0.00$\\
		$q^2 = (M_{\Lambda_b} - M_{\Lambda})^2$&$11.0\pm 0.00$&$4.04\pm 0.02$&$0.00\pm 0.13$&$11.5\pm 0.0$&$0.66\pm 0.07$&$-3.05\pm 0.11$\\
		\hline
	\end{tabular}
\end{table}

Finally, the helicity amplitudes are related to the form factors as~\cite{KornerSM}
\begin{eqnarray}\label{HelicityAmp}
&&a_\pm = \frac{G_F}{\sqrt{2}}{\cal C}_V f_V M_V\left[\sqrt{Q_-} \left(
f_1 \frac{M_+}{M_V} +f_2 \frac{M_V}{M_{\Lambda_b}}\right) \mp \sqrt{Q_+}\left(
g_1\frac{M_-}{M_V} - g_2 \frac{M_V}{M_{\Lambda_b}}
\right)\right]\,,\nonumber\\
&&b_\pm = \frac{G_F}{\sqrt{2}}{\cal C}_V f_V M_V\left[\sqrt{2 Q_-} \left(
-f_1  -f_2 \frac{M_+}{M_{\Lambda_b}}\right) \mp \sqrt{2 Q_+}\left(
-g_1+ g_2 \frac{M_-}{M_{\Lambda_b}}
\right)\right]\,,
\end{eqnarray}
where $M_\pm = M_{\Lambda_b} \pm M_\Lambda$ and $Q_\pm = M_\pm^2 - M_V^2$.
The decay widths and direct $CP$ asymmetries are given by
\begin{eqnarray}
&&\Gamma_{\Lambda_b \to \Lambda V}=
\frac{1}{16\pi} \frac{| \vec{p}_\Lambda|}{M_{\Lambda_b}^2}
\left(
|a_+|^2 + |a_-|^2 +|b_+|^2 +|b_-|^2 
\right)\,,
\nonumber\\
&&A_{CP}= \frac{\Gamma(\Lambda_b\to \Lambda\, V)-\Gamma(\overline{\Lambda}_b\to \overline{\Lambda}\, \overline{V})}{\Gamma(\Lambda_b\to \Lambda \,V)  +\Gamma(\overline{\Lambda}_b\to \overline{\Lambda}\, \overline{V})}\,.
\end{eqnarray}
The numerical results are shown in Table~\ref{BRandACP}, where the first and second uncertainties arise from the bag radius and CKM matrix elements, respectively. 
We see that with $N_c=3 $, $\Lambda_b \to \Lambda \omega$ is suppressed at  the level of $ 10^{-8}$ due to the cancellation of the effective Wilson coefficients, which is also found in the framework of the QCD factorization~\cite{Zhu:2018jet}. 
To take account the  NF effects, we treat $N_c$  as a parameter in the decay branching ratios, which are in general related to the decay processes.
Nonetheless, in this work, we will assume that $N_c$ is independent of the vector mesons.
 With the experimental branching ratio in $\Lambda_b \to \Lambda \phi$, we find that $N_c = 2.0\pm 0.3$, which is consistent with the $B$ meson decays.

In the factorization approach, 
$\alpha_{\lambda_1}$ and $\alpha_{\lambda_2}$ are independent of  ${\cal C}_V$ as can be seen from Eqs.~\eqref{alpha1}, \eqref{alpha2} and \eqref{HelicityAmp}. In addition, we find that in the bag model, they also depend little on the bag radius and the vector mesons. Explicitly, we have that 
\begin{equation}
\alpha_{\lambda_1} = -0.99\approx -1\,,~~~~~~\alpha_{\lambda_2} = 0.86\,.
\end{equation}
for
$\Lambda_b\to \Lambda V$ with $V=\{\phi,  \rho^0, \omega, K^{*0}\}$.
The values can be understood in the heavy quark limit, in which the light quark chiralities are related to the helicities. 
In Eq.~\eqref{Factorized}, the $s$ quark in $\Lambda$ is left-handed, and the  quark and anti-quark in $V$ have the same chiralities, resulting in $\lambda_1 = -1/2$ and $\lambda_2 = 0$.
We conclude that the amplitudes are dominated by $a_-$ in the framework of the factorization in the SM. 

On the other hand, once the NF contributions are considered, the arguments would not hold. In the $B$ meson decays with $b\to s$ transitions, we know that the NF contributions are mainly found in the negative helicities~\cite{hp1,hp2,hp3,hp4,hp5,hp6}, which lead to the so-called polarization puzzles in $B^0\to K^{*0}\phi$~\cite{Chen:2003jfa,BaBar:2003spf,Cheng:2001aa,Bauer:1986bm}. In analogy, we assume that the NF effects attribute solely to $b_+$  in the $b\to s$ transitions of $\Lambda_b$ decays.  As a result, since $b_+\approx 0$ in the factorizable amplitudes,  the branching ratios  increase along with the NF contributions, which meet well with the results in Table~\ref{BRandACP}.

\begin{table}\label{BRandACP}
	\caption{Branching ratios and direct CP asymmetries in units of $10^{-6}$ and $\%$ for $\Lambda_b \to \Lambda V$\,,   where
		the first and second uncertainties come from the bag radius and the CKM matrix elements in the SM, respectively, while the experimental branching ratio of $\Lambda_b \to \Lambda \phi$ is taken from Ref.~\cite{LambdabLambdaPhiExp}. }
	\begin{tabular}{l c|ccccc }
		\hline
		$V$&&$N_c$=1.7&$N_c$=2.0&$N_c$=2.3&$N_c$=3&Exp\\
		\hline
		\multirow{3}{*}{$\phi$}&$Br$&$ 6.90\pm0.65 \pm 0.30 $&$ 5.22\pm0.49\pm 0.24 $&$ 4.12\pm0.39 \pm 0.18 $&$ 2.67\pm0.25\pm 0.12 $& $ 5.18\pm 1.29 $\\
		&$A_{CP}$&$ 1.14\pm 0.00\pm 0.00 $&$ 1.16\pm 0.00\pm 0.00 $&$ 1.18\pm 0.00\pm 0.00  $&$ 1.21\pm 0.00 \pm 0.00 $\\
		&$\alpha_{\lambda_2}^N$&$-0.28 $&$ -0.05 $&$ 0.21 $&$ 0.86$\\
		\hline
		\multirow{3}{*}{$\rho^0$}&$Br$&$ 0.30\pm0.03\pm 0.02 $&$ 0.27\pm0.03\pm 0.02 $&$ 0.25\pm0.03 \pm 0.02$&$ 0.24\pm0.03 \pm 0.01$\\
		&$A_{CP}$&$ -2.15\pm 0.00\pm 0.04$&$ -1.61\pm 0.00\pm 0.01$&$ -1.11 \pm 0.00\pm 0.01$&$ -0.18\pm 0.00 \pm 0.00$\\
		&$\alpha_{\lambda_2}^N$&$0.43 $&$ 0.59 $&$ 0.71 $&$ 0.86$\\
		\hline
		\multirow{3}{*}{$\omega$}&$Br$&$ 2.98 \pm 0.35 \pm 0.15 $&$ 1.36\pm0.16 \pm 0.07 $&$ 0.57\pm 0.06 \pm 0.03$&$ 0.01\pm0.00 \pm 0.00$\\
		&$A_{CP}$&$ -1.88\pm 0.00\pm 0.07$&$ -1.87\pm 0.00\pm 0.07$&$ -1.86\pm 0.00\pm 0.07$&$ -1.55 \pm 0.00\pm 0.04$\\
		&$\alpha_{\lambda_2}^N$&$ -1 $&$  -0.99 $&$  -0.98 $&$ 0.86$\\
		\hline
		\hline
		\multirow{2}{*}{$K^{*0}$}&$Br$&$ 0.10\pm0.01\pm 0.00 $&$ 0.11\pm0.01\pm0.00 $&$ 0.12\pm0.02\pm 0.01 $&$ 0.14 \pm0.02 \pm 0.01 $\\
		&$A_{CP}$&$ -14.1\pm 0.0\pm 0.4 $&$ -13.5\pm 0.0\pm 0.3 $&$ -13.1 \pm 0.0\pm 0.3$&$ -12.6\pm 0.0\pm 0.5 $\\
		\hline
	\end{tabular}
\end{table}

In the $b\to s$ transitions, we assume that $a_-$ are totally factorizable, calculated with $N_c=3$, and $|b_+|^2$ can be obtained by subtracting the  $|a_-|^2$ contributions in the branching ratios. The numerical values of $\alpha_{\lambda_1}$  remain unaltered, since $|a_-|^2$ and $|b_+|^2$ share the same sign in Eq.~\eqref{alpha1}. In contrast, $\alpha_{\lambda_2}$ decrease along with $N_c$. The results are given in Table~\ref{BRandACP}. We see that $\alpha_{\lambda_2}$ are sensitive
to $N_c$. With $N_c = 2.0\pm 0.3$, we get that 
\begin{equation}
\alpha^N_{\lambda_2}= 0.0\pm 0.3 \,, \,\,-1\,, \,\, 0.6\pm 0.2
\end{equation}
for $\Lambda_b \to (\Lambda \phi, \Lambda \omega , \Lambda \rho^0)$, where the superscript of  ``$N$'' denotes  the  NF contribution in the scenario of the effective color number. Here, we find that $\Lambda_b \to \Lambda \omega$ is dictated by the NF contribution, which is consistent with those in
the literature~\cite{Zhu:2018jet,GeneralHsiao:2017tif}.


In the factorization approach,  the relative phases of $a_\pm$ and $b_\pm$ would vanish~(see Eq.~\eqref{HelicityAmp}), and therefore, one predicts that $\Delta_{t_i}^{(p)}=0$ for $i=1,2,3,4$. 
Although it only holds in the framework of the factorization,  with the scenario that the NF effects attribute to $b_+$ solely, $\Delta_{t_1}$
would  remain suppressed. Furthermore, $\Lambda_b \to \Lambda \phi$ is dominated by one weak phase, $V_{ts}^*V_{tb}$, so that  the effects of CP- and T-violation are highly suppressed. Explicitly, we have 
\begin{equation}\label{lasteq}
{\cal T}_{t_i}^{(p)}(\Lambda_b \to \Lambda \phi) \approx 0\,,~~~~~\text{for}~~i=1,2,3,4\,,
\end{equation}
which are independent of the factorization ansatz, providing a clean background to test the SM and search for new physics.

We now explore the possible contributions from new physics  to the T-violating observables.
Note that in $\Lambda_b \to \Lambda \omega/\rho^0$,  the $s$ quark is essentially left-handed in the SM, and thus
the experimental results with $\alpha_{\lambda_1}\ge 0 $ can be a smoking gun for new physics.
Particularly, $\Lambda_b \to \Lambda \omega$ acquires large contributions from right-handed penguin operators. 
To illustrate the effects, we concentrate on the possible effective Lagarangian from new physics, given by~\cite{ref3,last,Kagan:2004ia}
\begin{equation}\label{LNew}
{\cal L}^{\cal N}_{eff} = -\frac{G_F}{\sqrt{2}}\tilde{c}_5(\overline{s}_\alpha b_\alpha)_R \sum_q\left(
\overline{q}_\beta q_\beta
\right)_L\,,
\end{equation}
which contributes mainly to $a_+$ in the factorization approach, given as 
\begin{equation}
a^{\cal N}_+ = G_F\tilde{c}_5 f_\omega M_\omega\left[\sqrt{Q_-} \left(
f_1 \frac{M_+}{M_\omega} +f_2 \frac{M_\omega}{M_{\Lambda_b}}\right) + \sqrt{Q_+}\left(
g_1\frac{M_-}{M_\omega} - g_2 \frac{M_\omega}{M_{\Lambda_b}}
\right)\right]\,,
\end{equation}
where $\tilde{c}_5$ is the effective coefficient and  the superscript of ${\cal N}$ denotes new physics. With $\tilde{c}_5 > 10 ^{-3}$, new physics could potentially flip the sign of $\alpha_{\lambda_1}$.

On the other hand, 
due to the helicity conservation of the $s$ quark,
$(a_+\,, b_-)$ and $(a_-\,, b_+)$
 receive contributions from ${\cal L}_{eff}^{\cal N}$ and the SM,
respectively, and the CP-violating effects are suppressed due to the lack of the interferences. In contrast, ${\cal T}_{t_1}^p$,
depending on the complex phases between different helicities,
can be sizable. With the assumption of $|a_-|\approx |b_+|$ and $|a_+^{\cal N}| \approx |b_-^{\cal N}|$, we find that
\begin{equation}
{\cal T}_{t_1}^p \sim  \sqrt{ 1 -  \alpha_{\lambda_1} ^2 } \sin \phi_{\cal N}\,,
\end{equation}
which can be large when the phase of $\phi_{\cal N}$ from new physics is sizable. As a result, we conclude that the T-violating observables are useful in testing the complex phases for new physics.

\section{Conclusions}
We have parametrized the helicity amplitudes in terms of the angular distributions 
and  systematically studied the T-violating observables in $\Lambda_b \to \Lambda(\to p \pi^-) V(\to PP')$.
We have shown that all the relative complex phases among $a_\pm$ and $b_\pm$ can be interpreted as the T-odd correlations. By subtracting the effects 
from the FSIs, we have defined the true T-violating observables,
which could be measured in the experiments. 
In particular, we recommend the experiments on $\Delta_{t_1}^p$ and ${\cal T}^p_1$, which do not require $\Lambda_b$ to be polarized.
In addition, the polarization asymmetries of $\Lambda$ and $V$ have been defined
and their effects on the cascade decays have been  given.

The decays of $\Lambda_b \to \Lambda V$ in the SM have been examined  with the generalized factorization approach, which leads to the domination of $a_-$, resulting in that $\alpha_{\lambda_1}\approx -1$. Since the complex phases among $a_\pm$ and $b_\pm$ are identical, the factorization approach suggests that T-violating observables in the decays can not be observed. Nonetheless, the measured  branching ratio of $\Lambda_b\to \Lambda \phi$ indicates that the NF effects play an important role, resulting in that
the nonzero values of $\Delta_{t_i}^{(p)}$ do not necessary contradict to the SM.
However, as $\Lambda_b \to \Lambda \phi$ is dominated by a single weak phase, the true T-violating effects are not expected to be observed.
In Table \MakeUppercase{\romannumeral 4},  we have given
the branching ratios and direct CP asymmetries for the different values of the effective color number $N_c$.
We have found that $\alpha_{\lambda_2}$ depend heavily on the NF contributions.
Furthermore, the absolute value of $A_{CP}(\Lambda_b \to \Lambda K^{*0})$ has been expected to be larger than $10\%$.
We have also explored the possible effects from new physics. In particular, we have illustrated that
	the right-handed currents from new physics can potentially flip the sign of $\alpha_{\lambda_1}$ from negative to positive, resulting in a possible large T-violating effect.
 Finally, we recommend the future experiments on ${\cal T}_1^p(\Lambda_b \to \Lambda\phi)$ to test the SM and search for new physics.

\appendix
\section{T-transformation}
If the system respects the T symmetry, one has that~\cite{GroupTheory}
\begin{equation}
\langle f | U(\infty, -\infty) |i \rangle = \langle i^T | U(\infty, -\infty)| f ^T\rangle \,,
\end{equation}
for arbitrary initial and final states $|i\rangle$ and $| f\rangle$, respectively, where $U(t, t_0) $ is the time evolution operator from $t_0$ to $t$, and the superscript $T$ denotes the time-reversed state. Hence, in general,
the T-transformation relates $i\to f$ to $f^T\to i^T$ instead of $i^T \to f^T$.

In the first order of the weak interaction,
it is possible to relate the amplitudes between $i\to f$ and $i^T \to f^T$.
To do  this, we  adopt the interaction picture,
 in which the weak transition is described by ${\cal H}_{eff}$, and the unperturbed Hamiltonian corresponds to the strong interaction.   
 We have~\cite{GroupTheory}
\begin{equation}\label{A2}
\langle f;{\text{``out''}} | {\cal H}_{eff} | i \rangle  =  \langle i^T  |{\cal H}_{eff}^\dagger | f^T;{\text{``in''}} \rangle =  \langle i ^T |{\cal H}_{eff} | f^T;{\text{``in''}} \rangle  \,,
\end{equation}
where ``in'' and ``out''  denote $t\to \mp \infty$, respectively.
Here, the states are related as
\begin{equation}
U_0(\infty , -\infty ) |f; ``\text{in''}\rangle = |f; ``\text{out''}\rangle \,,
\end{equation}
where $U_0 $ represents the time evolution operator for the  unperturbed Hamiltonian~(strong interaction). 
In Eq.~\eqref{A2}, we have taken $|i\rangle$ as a single particle state of a stable hadron, having $|i;{\text{``in''}}\rangle = |i ; {\text{``out''}}\rangle $. Furthermore, in Eq.~\eqref{A2}, if $|f^T\rangle$ is an eigenstate of the FSI, we would have  $|f^T; {\text{``in''}}\rangle = e^{ic_f}|f^T; {\text{``out ''}}\rangle $ with $c_f$ the elastic rescattering phases, thereby leading to 
\begin{equation}\label{appenreal}
|\langle f;{\text{``out''}} | {\cal H}_{eff} | i \rangle  |^2=  |\langle i  |{\cal H}_{eff} | f^T;{\text{``out''}} \rangle  |^2 \,.
\end{equation}
As an application, for instance, after factorizing the pion decays,  $|f\rangle$ corresponds to the vacuum and
Eq.~\eqref{appenreal} demands the pion decay constants to be real.

\section*{ACKNOWLEDGMENTS}
We would like to thank Yi-Wen Lin for the assistance on the figure.

\end{document}